\documentclass[twoside,twocolumn,9pt]{article}
\usepackage{extsizes}
\usepackage[super,sort&compress,comma]{natbib} 
\usepackage[version=3]{mhchem}
\usepackage[left=1.5cm, right=1.5cm, top=1.785cm, bottom=2.0cm]{geometry}
\usepackage{balance}
\usepackage{times,mathptmx}
\usepackage{sectsty}
\usepackage{graphicx} 
\usepackage{lastpage}
\usepackage[format=plain,justification=justified,singlelinecheck=false,font={stretch=1.125,small,sf},labelfont=bf,labelsep=space]{caption}
\usepackage{float}
\usepackage{fancyhdr}
\usepackage{fnpos}
\usepackage[english]{babel}
\usepackage{array}
\usepackage{droidsans}
\usepackage{charter}
\usepackage[T1]{fontenc}
\usepackage[usenames,dvipsnames]{xcolor}
\usepackage{setspace}
\usepackage[compact]{titlesec}
\usepackage{hyperref}

\definecolor{cream}{RGB}{222,217,201}

\DeclareMathAlphabet{\mathcal}{OMS}{cmsy}{m}{n}
\usepackage{amsmath,amssymb}
\usepackage{mathrsfs}
\usepackage{subcaption}
\usepackage{mdwlist}
\usepackage{bm}
\usepackage{empheq}
\usepackage{verbatim}
\usepackage{bigints}
\usepackage{subcaption}
\usepackage{mathtools}
\usepackage{url}

\DeclareMathOperator*{\argmin}{\mathrm{argmin}}
\DeclareMathOperator{\tr}{\mathrm{tr}}
\DeclareMathOperator{\curl}{\mathrm{curl}}
\DeclareMathOperator{\dv}{\mathrm{div}}
\DeclareMathOperator{\grad}{\nabla}
\DeclareMathOperator{\epr}{\times}

\newcommand{\body}{\mathscr{B}}
\newcommand{\boundary}{\partial\body}
\newcommand{\area}{A}
\newcommand{\volume}{V}
\newcommand{\n}{\bm{n}}
\newcommand{\normal}{\bm{\nu}}

\newcommand{\e}{\bm{e}}
\newcommand{\alin}{a_\mathrm{lin}}
\newcommand{\aqua}{a_\mathrm{qua}}
\newcommand{\amin}{a_\mathrm{min}}

\newcommand{\acyl}{a_\mathrm{cyl}}
\newcommand{\almin}{\alpha_\mathrm{min}}

\newcommand{\grads}{\grad_{\!\mathrm{s}}}
\newcommand{\dd}{\operatorname{d}\!}
\newcommand{\dvs}{\dv_{\!\mathrm{s}}}

\newcommand{\free}{\mathcal{F}}
\newcommand{\Free}{\mathscr{F}}

\newcommand{\torframe}{(\e_r,\e_\varphi,\e_\psi)}
\newcommand{\torcoo}{(r,\varphi,\psi)}
\newcommand{\carframe}{(\e_x,\e_y,\e_z)}
\newcommand{\coner}{{C}^{1}(\mathbb{R})}
\newcommand{\conerd}{{C}^{h}(\mathbb{R}^{d})}

\newcommand{\cost}{\mathcal{C}}
\newcommand{\zero}{\bm{0}}
\newcommand{\nbend}{\n_\mathrm{b}}
\newcommand{\maxtwist}{\alpha_\mathrm{M}}
\newcommand{\rmt}{\sigma_\mathrm{rmt}}

\begin{document}

\pagestyle{fancy}
\thispagestyle{plain}
\fancypagestyle{plain}{

\renewcommand{\headrulewidth}{0pt}
}

\makeFNbottom
\makeatletter
\renewcommand\LARGE{\@setfontsize\LARGE{15pt}{17}}
\renewcommand\Large{\@setfontsize\Large{12pt}{14}}
\renewcommand\large{\@setfontsize\large{10pt}{12}}
\renewcommand\footnotesize{\@setfontsize\footnotesize{7pt}{10}}
\makeatother

\renewcommand{\thefootnote}{\fnsymbol{footnote}}
\renewcommand\footnoterule{\vspace*{1pt}%
\color{cream}\hrule width 3.5in height 0.4pt \color{black}\vspace*{5pt}} 
\setcounter{secnumdepth}{5}

\makeatletter 
\renewcommand\@biblabel[1]{#1}            
\renewcommand\@makefntext[1]%
{\noindent\makebox[0pt][r]{\@thefnmark\,}#1}
\makeatother 
\renewcommand{\figurename}{\small{Fig.}~}
\sectionfont{\sffamily\Large}
\subsectionfont{\normalsize}
\subsubsectionfont{\bf}
\setstretch{1.125} 
\setlength{\skip\footins}{0.8cm}
\setlength{\footnotesep}{0.25cm}
\setlength{\jot}{10pt}
\titlespacing*{\section}{0pt}{4pt}{4pt}
\titlespacing*{\subsection}{0pt}{15pt}{1pt}

\fancyfoot{}
\fancyfoot[RO]{\footnotesize{\sffamily{1--\pageref{LastPage} ~\textbar  \hspace{2pt}\thepage}}}
\fancyfoot[LE]{\footnotesize{\sffamily{\thepage~\textbar\hspace{3.45cm} 1--\pageref{LastPage}}}}
\fancyhead{}
\renewcommand{\headrulewidth}{0pt} 
\renewcommand{\footrulewidth}{0pt}
\setlength{\arrayrulewidth}{1pt}
\setlength{\columnsep}{6.5mm}
\setlength\bibsep{1pt}

\makeatletter 
\newlength{\figrulesep} 
\setlength{\figrulesep}{0.5\textfloatsep} 

\newcommand{\topfigrule}{\vspace*{-1pt}%
\noindent{\color{cream}\rule[-\figrulesep]{\columnwidth}{1.5pt}} }

\newcommand{\botfigrule}{\vspace*{-2pt}%
\noindent{\color{cream}\rule[\figrulesep]{\columnwidth}{1.5pt}} }

\newcommand{\dblfigrule}{\vspace*{-1pt}%
\noindent{\color{cream}\rule[-\figrulesep]{\textwidth}{1.5pt}} }

\makeatother

\twocolumn[
  \begin{@twocolumnfalse}
\vspace{3cm}
\sffamily
\begin{tabular}{m{4.5cm} p{13.5cm} }

 & \noindent\LARGE{\textbf{Lily-like twist distribution in toroidal nematics$^{\ddag}$}} \\
\vspace{0.3cm} & \vspace{0.3cm} \\

 & \noindent\large{Andrea Pedrini,\textit{$^{a}$} Marco Piastra,\textit{$^{b}$} and Epifanio G. Virga\textit{$^{c\ast}$}} \\
\vspace{0.3cm}\\
& \noindent\normalsize{Toroidal nematics are droplets of nematic liquid crystals in the form of a circular torus. When the nematic director is subject to planar degenerate boundary conditions, the bend-only director field with vector lines along the parallels of all internal torodial shells is an equilibrium solution for all values of the elastic constants. Local stability analyses have shown that an instability is expected to occur for sufficiently small values of the twist elastic constant. It is natural to conjecture that in this regime the global equilibrium would be characterized by a maximum twist deflection on the boundary of the torus, with a twist distribution over the torus' cross-section represented by a \emph{fennel-like} surface. We prove that surprisingly the stable twist distribution is instead  represented by a \emph{lily-like} surface. Thus the overall maximum twist deflection falls well within the torus. To cope with the complexity of the elastic free-energy functional in the fully non-linear setting, we developed an \emph{ad hoc} deep-learning optimization method, which here is also duly validated and documented for it promises to be applicable to other similar problems, equally intractable analytically.}

\end{tabular}
 \end{@twocolumnfalse} \vspace{0.6cm}
]

\renewcommand*\rmdefault{bch}\normalfont\upshape
\rmfamily
\section*{}
\vspace{-1cm}


\footnotetext{\textit{$^{a}$~Dipartimento di Matematica, Universit\`a di Pavia, via Ferrata 5, 27100 Pavia, Italy. E-mail: andrea.pedrini@unipv.it}}
\footnotetext{\textit{$^{b}$~Dipartimento di Ingegneria Industriale e dell'Informazione, Universit\`a di Pavia, via Ferrata 5, 27100 Pavia, Italy. E-mail: marco.piastra@unipv.it}}
\footnotetext{\textit{$^{c}$~Dipartimento di Matematica, Universit\`a di Pavia, via Ferrata 5, 27100 Pavia, Italy.  E-mail: eg.virga@unipv.it}}


\footnotetext{\ddag~All authors have contributed equally to this work.}

\section{Introduction}\label{sec:intro}
Toroidal nematics are typically droplets of nematic liquid crystals in the shape of a circular torus surrounded by an isotropic fluid that enforces degenerate planar anchoring conditions for the nematic director $\n$. We imagine that the droplet's shape $\body$ is somehow prescribed and inquire about the most stable equilibrium nematic structure inside it. This problem was  studied by Koning et al.\cite{koning:saddle-splay} who envisaged a first destabilization mechanism by twist-injection from the surface into  a bend-only nematic texture.

The elastic free energy stored in $\body$ is given by the functional
\begin{equation}\label{eq:Frank_free_energy}
\begin{split}
\Free[\bm{n}] = \int_{\body}\Big\{&
\frac{1}{2}\Big[K_{1}(\dv\bm{n})^{2} + K_{2}(\bm{n}\cdot\curl\bm{n})^{2} + K_{3}|\bm{n}\epr\curl\bm{n}|^{2}\Big]\\&+
K_{24}\Big[\tr(\grad\bm{n})^{2}-(\dv\bm{n})^{2}\Big]\Big\}\dd\volume\,,
\end{split}
\end{equation}
where $\dd\volume$ is the volume element and $K_1$, $K_2$, $K_3$, and $K_{24}$ are Frank's elastic constants, weighting the costs of splay, twist, bend, and saddle-splay distortions, respectively (as explained in a number of textbooks,~\cite{degennes:physics,virga:variational,stewart:static} which the reader is referred to for further details).
Frank's energy functional $\Free$ in \eqref{eq:Frank_free_energy} measures the elastic cost involved in distorting a uniform state where $\nabla\n\equiv\zero$. As first remarked by
Ericksen,~\cite{ericksen:inequalities} this functional, which vanishes on any uniform state, is  positive semi-definite only if the elastic constants obey the inequalities 
\begin{equation}\label{eq:Ericksen_inequalities}
K_{1}\geqslant K_{24}\geqslant0,\quad K_{2}\geqslant K_{24},\quad K_{3}\geqslant 0\,,
\end{equation}
which are also referred to as Ericksen's inequalities. Here, for convenience, we
shall scale all elastic constants to $K_2$, assumed to be positive,
\begin{equation}\label{eq:scaled_elastic_constants}
k_3:=\frac{K_3}{K_2}\quad\text{and}\quad k_{24}:=\frac{K_{24}}{K_2},
\end{equation}
so that Ericksen's inequalities become
\begin{equation}\label{eq:rescaled_Ericksen_inequalities}
k_3\geqslant0\quad\text{and}\quad 0\leqslant k_{24}\leqslant1.
\end{equation}

As is well known, the $K_{24}$ energy in \eqref{eq:Frank_free_energy} is a null-Lagrangian and can be transformed into a surface energy, so that $\Free$ takes the equivalent form
\begin{equation}\label{eq:frank}
\begin{split}
\Free[\bm{n}] = &\frac{1}{2} \int_{\body}\Big[K_{1}(\dv\bm{n})^{2} + K_{2}(\bm{n}\cdot\curl\bm{n})^{2} + K_{3}|(\grad\bm{n})\bm{n}|^{2}\Big]\dd\volume \\
&+  \int_{\boundary}K_{24}\big[(\grads\bm{n})\bm{n}-(\dvs\bm{n})\bm{n}\big]\cdot\bm{\nu}\dd\area\,,
\end{split}
\end{equation}
where $\dd\area$ is the area element and $\bm{\nu}$ is the outer unit normal to $\boundary$.

We want to find the director field $\n$ that minimizes $\Free$ subject to the boundary condition
\begin{equation}\label{eq:constraint}
\n\cdot\normal\equiv0\quad\text{on}\quad\boundary\,.
\end{equation}
As stated, this is a formidable variational problem. Luckily, a number of  partial results are known from the literature. First, there is a special, bend-only director field $\nbend$ whose vector lines are parallel circles with centers on the straight axis of the torus, also called the \emph{parallels} of $\body$. This field is a \emph{universal solution}\cite{ericksen:general,marris:universal,marris:addition} for the Euler-Lagrange equation associated with $\Free$, for any choice of the elastic constants. This means that $\nbend$, which complies with \eqref{eq:constraint}, is also an equilibrium configuration for our variational problem. It remains to be seen whether $\nbend$ is the absolute minimizer of $\Free$ as well.

The local stability of $\nbend$ for toroidal nematics was first studied by Koning et al.\cite{koning:saddle-splay} and then further refined.\cite{pedrini:instability} In brief, for any given $k_{24}$, $\nbend$ becomes unstable for $k_3$ sufficiently large, that is, when the cost of bending is sufficiently greater than the cost of twisting. The critical values of $k_3$ increases with decreasing $k_{24}$, thus suggesting that this is a surface-dominated transition, as \eqref{eq:frank} identifies $K_{24}$ as a surface-like elastic constant. When $\nbend$ becomes unstable a certain degree of twist is likely to be injected into the torus, but nothing is known about how the newly acquired twist is distributed inside the torus: the local stability analysis cannot answer this question. One other thing we know from symmetry: the chiral degeneracy of the system entails that the energy minimizers will come in pairs, with equal energy but opposite chirality.\cite{koning:saddle-splay,pedrini:instability}

In this paper, we solve the fully non-linear problem and find the optimal twist distribution in toroidal nematics when $\nbend$ is unstable. If the interpretation of $\nbend$ as being surface-driven were to be supported by the fully non-linear results, one would expect the minimizers of $\Free$ to exhibit the maximum twist deflection $\alpha_\mathrm{M}$ at the boundary $\boundary$ of the torus. On the contrary, we shall  see
that this is \emph{not} the case, as $\maxtwist$ falls generally in the interior of $\body$: more precisely, on every circular cross-section of $\body$, the twist angle distribution has typically a \emph{lily-like} appearance, as opposed to the \emph{fennel-like} surface that would describe the twist distribution had $\maxtwist$ fallen on $\boundary$. 

In other words, in toroidal nematics, the twist favoured by surface elasticity, does not extend freely up to the boundary, as it would happen in a cell bounded by planar plates, but it remains self-confined in the interior, giving rise to a characteristic twist structure which has the potential of being detectable optically. No doubt all this is conjured by the toroidal geometry, whether it is typical of that we cannot tell.

The paper is organized in the following way. Section~\ref{sec:reduced_energy} illustrates the special class of nematic director fields $\n$ within which we minimize $\Free$; in this class, which includes $\nbend$ as well as  quite different fields, we write a \emph{reduced} energy functional. In Section~\ref{sec:deep_learning}, we summarize the deep-learning method applied to this minimization problem, highlighting the mathematical infrastructure behind the code that we devised for this problem. In Section~\ref{sec:results}, we illustrate the results obtained through our method and comment on their physical significance, pausing on the details of the generic twist structure characterizing the states with minimum energy. Section~\ref{sec:validation} is devoted to a number of validation tests for our computations, both numerical and analytical in character. In particular, we determine numerically the curve in parameter space marking  the instability onset for the bend-only solution $\nbend$ and contrast it against the known analytical estimates.\cite{koning:saddle-splay,pedrini:instability} Finally, in Section~\ref{sec:conclusions}, we draw the conclusions of this work and we discuss possible extensions to other soft matter systems of the deep-learning method adopted here.  The paper is closed by an Appendix, where we outline the features of the accompanying code.

\section{Reduced energy functional}\label{sec:reduced_energy}
\begin{figure}[h]
 \centering
 \includegraphics[width=\linewidth]{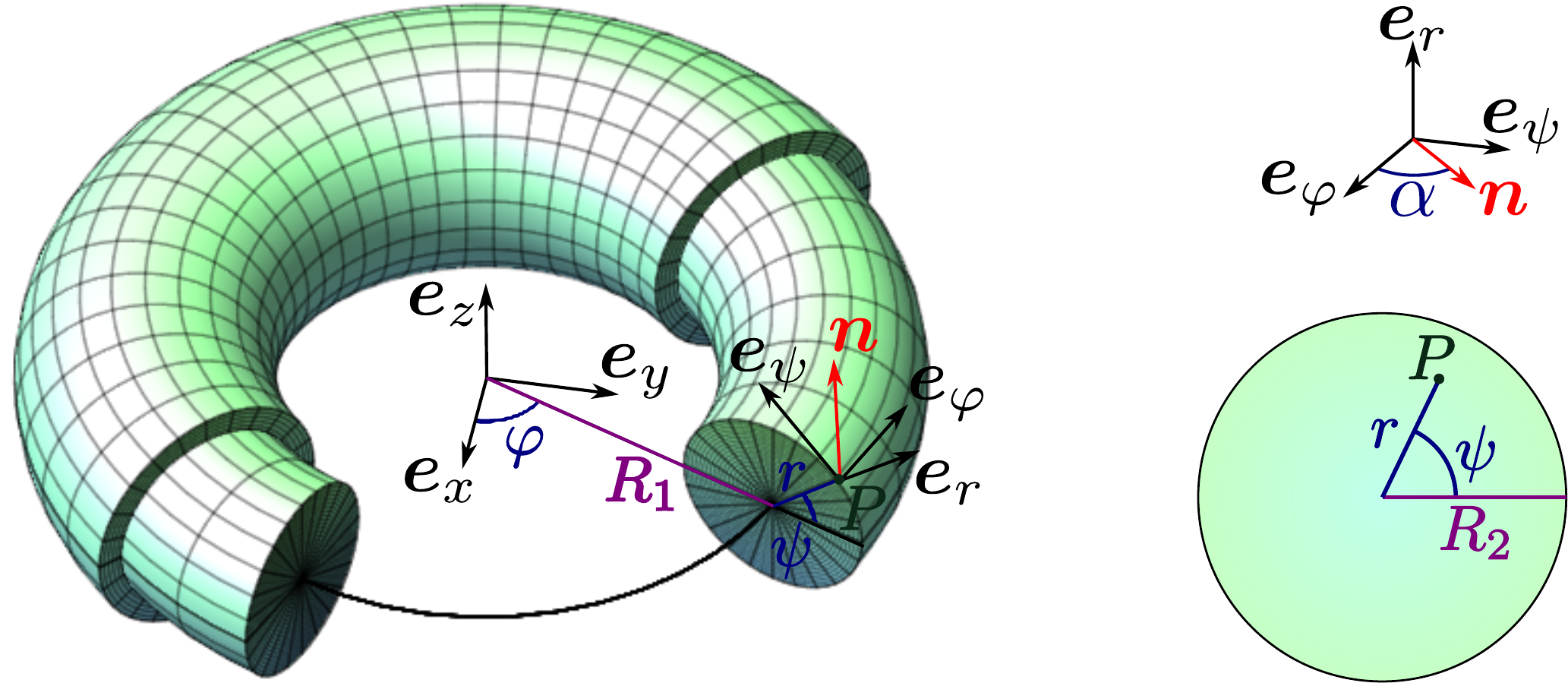}
  \caption{(Colour online) Representation of a circular torus $\body$ (left) obtained by revolving a disk of radius $R_{2}$ (bottom-right) around the $z$-axis. The centreline of $\body$ has radius $R_{1}$. Each point $P$ of the torus is described by the toroidal coordinates $\torcoo$ and the nematic vector field $\n$ at $P$ is represented in the associated toroidal frame $\torframe$ (top-right) through the angle $\alpha$, as in Equation~\eqref{eq:n_alpha_representation}.}
  \label{fig:torus}
\end{figure}

A toroidal nematic is described as a circular torus $\body$. We denote by $R_1$ and $R_2<R_1$ the radius of the inner circular axis of $\body$ and the radius of its meridian cross-sections (see Fig.~\ref{fig:torus}). We employ \emph{toroidal coordinates} $\torcoo$ to describe points in $\body$: $(\psi,r)$ are polar coordinates in the meridian cross-section of $\body$ with a plane making the angle $\varphi$ with a given reference plane, taken as the $(x,z)$ plane of a Cartesian frame $\carframe$ (see Fig.~\ref{fig:torus}). Thus $r$ ranges in the interval $[0,R_2]$ and both $\varphi$ and $\psi$ in $[0,2\pi)$. The coordinates $\torcoo$ are orthogonal and the associated frame $\torframe$ is positively oriented, that is, $\e_\psi=\e_r\times\e_\varphi$. In these coordinates, the volume and area elements are given by
\begin{equation}\label{eq:volume_area_elements}
\dd V=r(R_1+r\cos\psi)\dd r\dd\varphi\dd\psi,\quad\dd A=r(R_1+r\cos\psi)\dd\varphi\dd\psi\,.
\end{equation}

The special class of director fields that we shall consider in $\body$ are represented by
\begin{equation}\label{eq:n_alpha_representation}
\n=\cos\alpha\e_\varphi+\sin\alpha\e_\psi\,,
\end{equation}
where $\alpha=\alpha(r,\psi)$. These fields are characterized by being independent of the azimuth $\varphi$ and having zero radial component. Moreover, we shall assume that $\alpha(0,\psi)\equiv0$ to avoid a \emph{disclination} of $\n$ along the centreline of $\body$ (which would cost infinite energy in Frank's model). It follows from \eqref{eq:n_alpha_representation} that $\n=\nbend$ for $\alpha\equiv0$ and that reversing the sign of $\alpha$ reverses the \emph{chirality} of $\n$, that is, its sense of \emph{winding} around the torus.\cite{koning:saddle-splay}

To make our variables dimensionless, we shall scale lengths to $R_2$, so that
\begin{equation}\label{eq:sigma_definition}
\sigma:=\frac{r}{R_2}\in[0,1]\,,
\end{equation}
and we shall call $\eta\in[0,1]$ the ratio
\begin{equation}\label{eq:eta_definition}
\eta:=\frac{R_2}{R_1}\,.
\end{equation}

Following Koning et al.~\cite{koning:saddle-splay}, we further restrict the admissible class of director fields by
requiring that $\dv\n=0$, which amounts to the following differential equation for $\alpha$,
\begin{equation}\label{eq:divergence_free_condition}
\alpha_{,\psi} = \frac{\eta\sigma\sin\psi\tan\alpha}{1+\eta\sigma\cos\psi}\,,
\end{equation}
which can be easily integrated, delivering
\begin{equation}\label{eq:divergence_free_alpha}
 \alpha=\arcsin\frac{a(\sigma)}{1+\eta\sigma\cos\psi}\,,
\end{equation}
where $a(\sigma)$ is a real function of $\sigma$ only. For \eqref{eq:divergence_free_alpha} to obey the inequality
$|\sin\alpha|\leq1$, $a$ must be such that  $|a(\sigma)|\leq 1+\eta\sigma\cos\psi$ for all $\psi\in[0,2\pi]$, which is satisfied whenever
\begin{equation}
\label{eq:barrierscond}
 |a(\sigma)|\leq1-\eta\sigma\,.
\end{equation}

Thus, eventually, the energy functional $\Free[\n]$ can be reduced to a functional $\free[a]$ in the single scalar function $a$ subject to \eqref{eq:barrierscond} and
\begin{equation}\label{eq:a(0)=0}
a(0)=0\,.
\end{equation}
Making repeated use of Appendices A2 and A3 of Pedrini and Virga\cite{pedrini:instability}, and exercising patience, we finally arrive from \eqref{eq:frank} to the following formula for $\free$,

\begin{eqnarray}\label{eq:energy_functional}
 %
& \dfrac{\mathcal{F}[a]}{2\pi^{2}R_{1}K_{2}} = \displaystyle{\int_{0}^{1}}\bigg\{ \bigg(\frac{1}{C(a(\sigma))} + \frac{1}{C(-a(\sigma))}\bigg)\frac{\sigma}{2} a^{2}_{,\sigma}(\sigma) \nonumber\\
 &\qquad\qquad+\bigg(\frac{1}{C(a(\sigma))} - \frac{1}{C(-a(\sigma))}\bigg)  a_{,\sigma}(\sigma)\nonumber\\
  &\qquad\qquad- \bigg(\frac{1}{C(a(\sigma))} + \frac{1}{C(-a(\sigma))}\bigg)  a(\sigma)a_{,\sigma}(\sigma)\nonumber\\
 &\qquad\qquad + \frac{(1-a(\sigma))^{2}}{2\sigma C(a(\sigma))}+ \frac{(1+a(\sigma))^{2}}{2\sigma C(-a(\sigma))}+ \frac{2a(\sigma)a_{,\sigma}(\sigma)}{(1-\eta^{2}\sigma^{2})^{\frac{3}{2}}} \nonumber\\
 &\qquad\qquad -\frac{ 1}{\sigma(1-\eta^{2}\sigma^{2})^{\frac{1}{2}}}+ \frac{2+7\eta^{2}\sigma^{2}}{2\sigma(1-\eta^{2}\sigma^{2})^{\frac{5}{2}}}a^{2}(\sigma)\nonumber\\
 &\qquad\qquad - \frac{3\eta^{4}\sigma^{4}+24\eta^{2}\sigma^{2}+8}{8\sigma(1-\eta^{2}\sigma^{2})^{\frac{9}{2}}}a^{4}(\sigma)\nonumber \\
 &\qquad+ k_{3} \bigg[\frac{3\eta^{4}\sigma^{4}+24\eta^{2}\sigma^{2}+8}{8\sigma(1-\eta^{2}\sigma^{2})^{\frac{9}{2}}}a^{4}(\sigma) \nonumber \\
 &\qquad\qquad + \frac{1}{\sigma(1-\eta^{2}\sigma^{2})^{\frac{1}{2}}} - \frac{C(a(\sigma)) + C(-a(\sigma))}{2\sigma} \nonumber\\
 &\qquad\qquad -\frac{3\eta^{2}\sigma}{(1-\eta^{2}\sigma^{2})^{\frac{5}{2}}}a^{2}(\sigma)\bigg]\bigg\}\dd\sigma - 2k_{24}\frac{a^{2}(1)}{(1-\eta^{2})^{\frac{3}{2}}}\,,
\end{eqnarray}
where $C(a(\sigma)) = \sqrt{\left(1-a(\sigma)\right)^{2}-\eta^{2}\sigma^{2}}$.

Clearly, $\free[a]=\free[-a]$, showing that minimizers of $\free$ come in pairs of functions differing in sign. We shall systematically consider only one member of such a pair, keeping in mind that it also represents its companion with opposite sign (and chirality).

The (dimensionless) energy of $\nbend$ is readily obtained from \eqref{eq:energy_functional} as
\begin{equation}\label{eq:energy_parallels}
 \dfrac{\mathcal{F}[0]}{2\pi^{2}R_{1}K_{2}} = k_{3}\int_{0}^{1} \frac{\eta^{2}\sigma^{2}}{\sigma(1-\eta^{2}\sigma^{2})^{\frac{1}{2}}} \dd\sigma =  k_{3}\bigg(1-\sqrt{1-\eta^{2}}\bigg)\,.
\end{equation}
The excess energy associated with a generic $a$ will then be given by
\begin{equation}\label{eq:energy_difference}
 \mathcal{E}[a]:=\dfrac{\mathcal{F}[a]-\mathcal{F}[0]}{2\pi^{2}R_{1}K_{2}}\,.
\end{equation}
Formula \eqref{eq:energy_functional} has the advantage of being valid for all director fields in the class \eqref{eq:n_alpha_representation}, no matter how they differ from $\nbend$, but it is intractable analytically. We devised a numerical scheme to minimize $\free[a]$, which is outlined in the following section.

\section{Deep-learning method}\label{sec:deep_learning}
In general, deep-learning methods are constructed to find  smooth functions $y^\ast$ (typically, of class $\coner$) that minimize a given \emph{cost} functional $\cost[y]$.\footnote[3]{All definitions and results presented in this section are more generally valid  for all spaces $\conerd$.} Formally, we shall write that
\begin{equation}\label{eq:minimisation_problem}
 y^{*} := \argmin_{y\in\coner}\cost[y].
\end{equation}
In general, $\argmin_{y\in\coner}\cost[y]$ denotes the \emph{set} of functions in $\coner$ where $\cost$ attains its minimum. Here, however, we shall adopt the simplifying assumption (corroborated  by physical intuition) that, possibly apart from a geometric degeneracy (such as the sign parity of both $\mathcal{F}$ and $\mathcal{E}$), $\cost$ has a \emph{unique} minimizer. 

Deep-learning methods are based on a \emph{nested} (or \emph{deep}) representation of $y$ such as
\begin{equation}\label{eq:representation}
 y^{(n)}(x) := \bm{w} \cdot \tanh(\bm{W}_{k} \tanh( {\cdots} \bm{W}_{2}\tanh(\bm{w}_{1}x+\bm{b}_{1})+\bm{b}_{2}\cdots) + \bm{b}_{k}) \,, 
\end{equation}
 where $\bm{w}$, $\bm{b}_{1},\bm{b}_{2},\dots,\bm{b}_{k}$ are vectors and $\bm{W}_2,\dots,\bm{W}_k$ are matrices, collectively represented by the set of parameters
$\vartheta^{(n)} := \{\bm{w},\bm{w}_{1},\bm{W}_{2},\dots,\bm{W}_{k},\bm{b}_{1},\bm{b}_{2},\dots,\bm{b}_{k}\}$. Here $n$ is the overall number of scalars contained in $\vartheta^{(n)}$ and $k$ is the \emph{depth} of the representation. The function $\mathrm{tanh}$ is applied element-wise to its vector and matrix arguments.\cite{goodfellow:deep}

Representation \eqref{eq:representation}, which is also called \emph{deep neural network} (DNN), is often said to be \emph{universal}. Informally, this means that \eqref{eq:representation} is a versatile and adaptable representation of a function through a finite number of scalar parameters. More technically, it has been proved\cite{hornik:approximation} that for all $y\in\coner$, all compact subset $X$ of $\mathbb{R}$, and all $\varepsilon>0$, there exist $n$ and $\vartheta^{(n)}$ such that
\begin{equation}
 \|y-y^{(n)}\|_{1,X}<\varepsilon,
\end{equation}
where $\|y\|_{1,X}:=\max(\sup_{x\in X}y(x),\sup_{x\in X}y_{,x}(x))$ and $y_{,x}$ denotes the derivative of $y$.

Adopting the above representation, we change the problem of  minimizing $\cost$ over $\coner$ into that of finding   
\begin{equation}
 \vartheta^{(n)*} := \argmin_{\vartheta^{(n)}}\cost^{(m)}[y^{(n)}]\,,
\end{equation}
where, to ease the numerics, the cost functional $\cost^{(m)}$ is $\cost$ evaluated on a discretised function $y^{(n)}$ defined over a finite dataset $\{x^{(i)}\in\mathbb{R}\}_{i=0}^{m}$ of input points. Thus, we reduce a minimization 
problem in a function space into an optimization problem in a finite, $n$-dimensional parameter space. Clearly, by construction, this method is only capable of identifying in an approximate fashion the smooth minimizers of the cost functional $\cost$; these latter are assumed both to exist and to bear physical meaning.

The deep-learning method outlined above is here applied to the problem at hand by taking $\mathcal{E}$ in \eqref{eq:energy_difference} as $\cost$. Accordingly, $\mathcal{E}[a]$ is replaced by the discretized version
 $\mathcal{E}^{(m)}[a^{(n)}]$, where $a^{(n)}$ is subject to the constraint
 \begin{equation} \label{eq:a_constraint}
a^{(n)}(0)=0
 \end{equation}
 and evaluated over 
the dataset $\{\sigma^{(i)}\in[0,1]\}_{i=0}^{m}$, with $\sigma^{(0)}=0$ and $\sigma^{(m)}=1$. The optimization problem then becomes: find
\begin{equation}\label{eq:optimization_problem}
 \vartheta^{(n)*} := \argmin_{\vartheta^{(n)}}\mathcal{E}^{(m)}[a^{(n)}],
\end{equation}
where
\begin{equation}\label{eq:minimizer_boundary}
 a^{(n)}(\sigma) := y^{(n)}(\sigma) - y^{(n)}(0)\,,
\end{equation}
so that the boundary condition \eqref{eq:a_constraint} is automatically satisfied.

We leave out a number of interesting mathematical questions, such as the degree of regularity of the minimizers of $\mathcal{E}[a]$ and the convergence of both $\mathcal{E}^{(m)}[a^{(n)}]$ and $a^{(n)}$ as both $m$ and $n$ are increased. These issues, though neither trivial nor irrelevant, exceed the scope of our present work, which is more concerned with the physical relevance of the results and treats the mathematical tools employed here as \emph{learning experiments} for an automatic resolution of our problem. This is why our code is made available in a workable fashion alongside this paper, see the Appendix for further explanatory notes. Here we are content to give some additional details on how the code was implemented to arrive at the results described in the following section.

\subsection{Numerical optimization: $\text{TensorFlow}^{\circledR}$}
TensorFlow\footnote{TensorFlow, the TensorFlow logo and any related marks are trademarks of Google Inc.} is a recent software tool\cite{abadi:tensorflow_large-scale,abadi:tensorflow_system} for
the numerical solution to minimization problems such as \eqref{eq:optimization_problem}. 
In particular, TensorFlow allows translating functionals like $\mathcal{E}^{(m)}[a^{(n)}]$
 into a \emph{flow graph} in which each node represents an elementary operation and each edge represents a dependency of an operator on its operands. In addition, TensorFlow allows computing the partial derivative of a flow graph with respect to any of its nodes via a technique called \emph{automatic differentiation}. \cite{bartholomew-biggs:automatic} The result of any such differentiation is yet another flow graph that \emph{shares} the relevant parts of the flow graph being derived. In passing, we note that the derivative $a_{,\sigma}$ in Equation~\eqref{eq:energy_functional} was also computed in this way.

Flow graphs can also be used to compute the numerical values of functions and functionals by specializing variables and parameters to definite tensorial values. With TensorFlow, such computations can be carried on transparently on either standard CPUs or GPU-accelerated hardware.

The optimization process that, in the case at hand, aims to obtain minimizing values of the parameters $\vartheta^{(n)}$ is performed by applying specific optimization operators, also provided by TensorFlow, which simplify the process of taking derivatives with respect to $\vartheta^{(n)}$ and applying iterative, quasi-second order methods.\cite{ruder:overwview} Processes of the above kind can only lead to \emph{local} minimizers for $\vartheta^{(n)}$ hence are dependent on the assigned initial values.

TensorFlow is an open source software package that runs on several different hardware platforms and operating systems.

\subsection{Experiments}
We adopted an approximator $a^{(n)}$ with $k=5$ and $\bm{w}, \bm{w}_{1}, \bm{b}_{1}, \dots, \bm{b}_{5}\in\mathbb{R}^{10}$, $\bm{W}_{2}, \dots, \bm{W}_{5}\in\mathbb{R}^{10}\times\mathbb{R}^{10}$. The discretized functional $\mathcal{E}^{(m)}$ was obtained by  using a dataset with $m=1000$ and $\sigma^{(i)} = {i}/{m}$. Among several possible choices, we selected Adam\cite{kingma:adam} as the quasi-second order method needed for the optimization. At the beginning of each run $\vartheta^{(n)}$ was assigned small random values. For deterministic repeatability, specific \emph{seeding} policies were adopted for pseudo-random number generators. The most effective values for the \emph{hyperparameters} $n$, $k$, and $m$ were determined experimentally.

All our experiments were carried out on a workstation $\text{Intel}^{\circledR}$ Xeon(R) CPU E3-1240 v3 @ 3.40GHz x 8 with 16 GB of RAM, equipped with Quadro K2200 GPU with 4 GB of DRAM.

In the rest of the paper, $\amin$ will always denote  the (presumably unique) minimizer found as a result of the numerical optimization process.

\section{Results and discussion}\label{sec:results}
We performed many different optimizations of $\mathcal{E}$ for several choices of the ratio $\eta$ and the normalized elastic constants $k_{3}$ and $k_{24}$; we illustrate here only a few representative experiments to highlight the main properties of the minimizers $\amin$. 

\begin{figure}[h]
 \centering
 \includegraphics[width=0.8\linewidth]{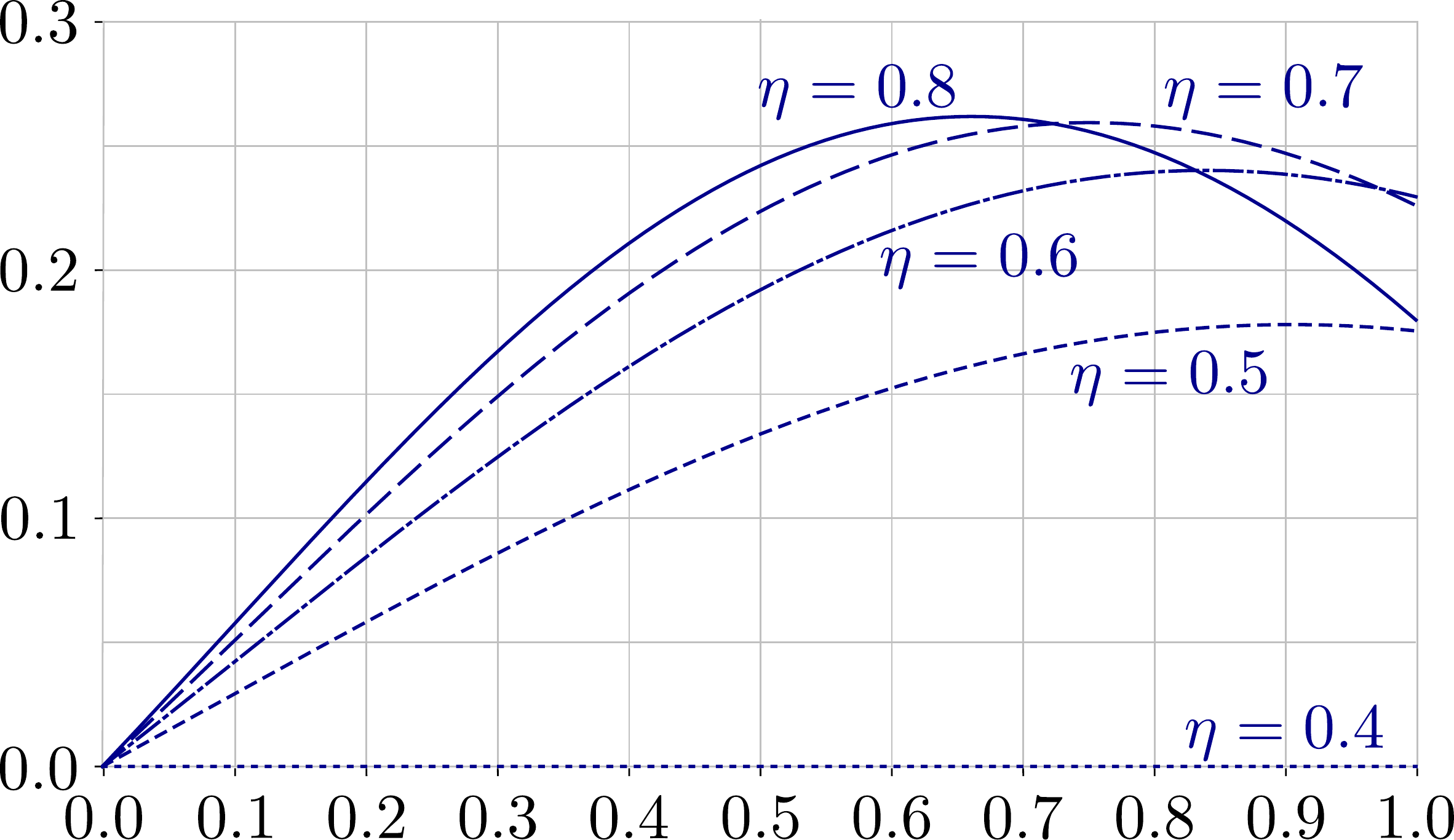}
 \caption{(Colour online) The minimizer $\amin$ of $\mathcal{E}$ for $k_{3}=7$, $k_{24}=0.5$, and different choices of $\eta$.}
 \label{fig:a_eta}
\end{figure}

\begin{figure}[h]
 \centering
 \includegraphics[width=0.8\linewidth]{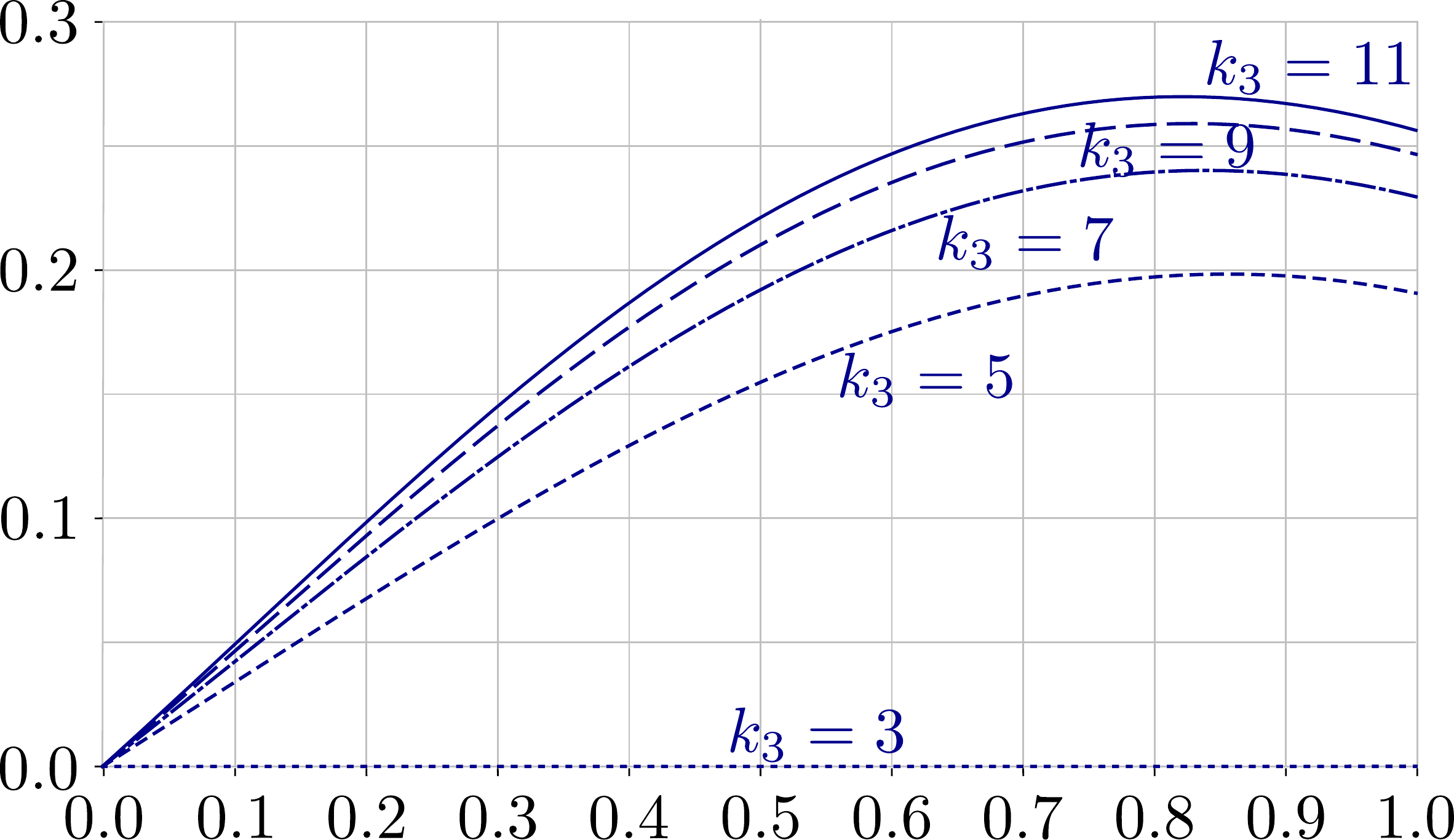}
 \caption{(Colour online) The minimizer $\amin$ of $\mathcal{E}$ for $\eta=0.6$, $k_{24}=0.5$, and different choices of $k_{3}$.}
 \label{fig:a_k3}
\end{figure}

\begin{figure}[h]
 \centering
 \includegraphics[width=0.8\linewidth]{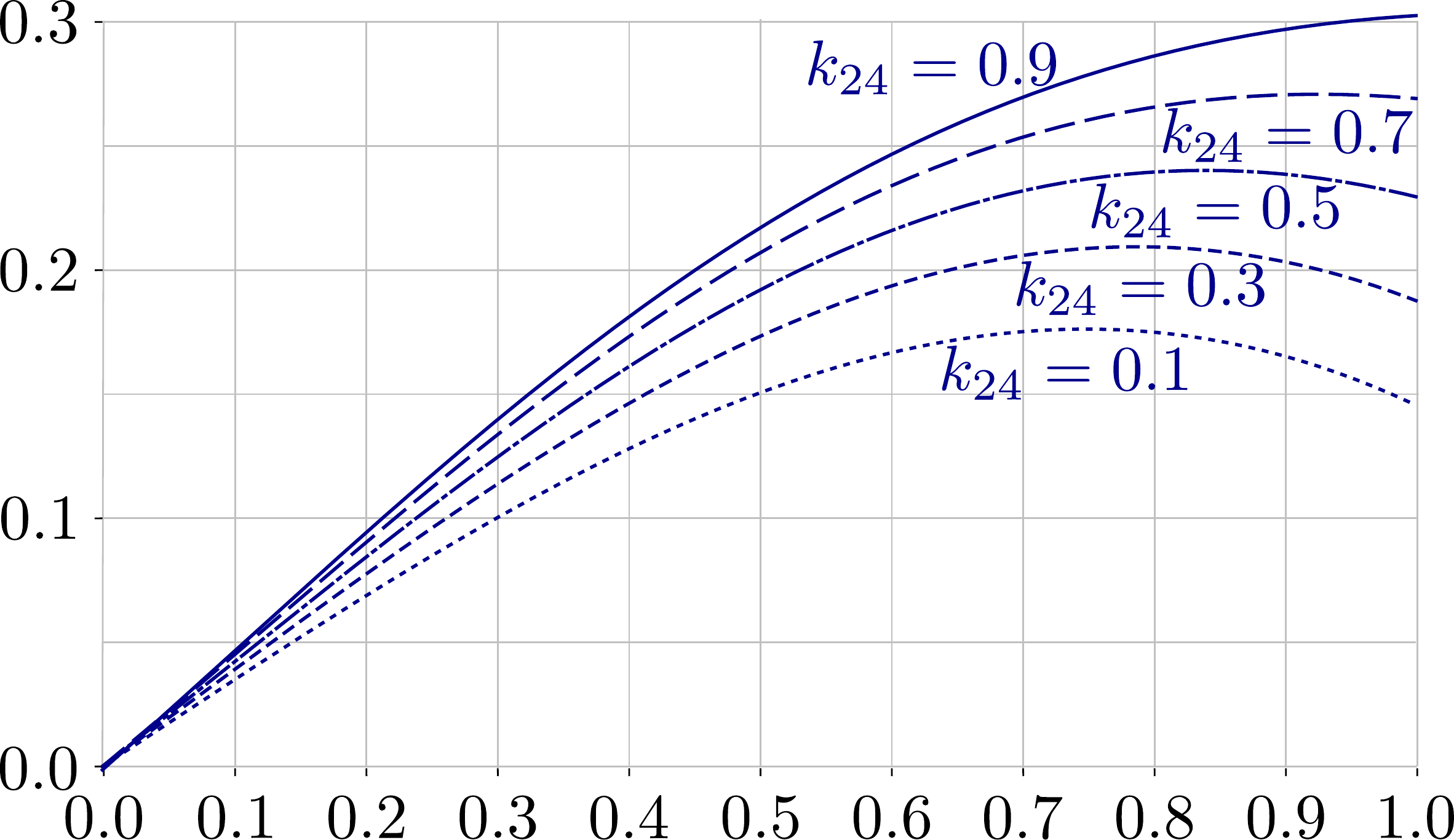}
 \caption{(Colour online) The minimizer $\amin$ of $\mathcal{E}$ for $\eta=0.6$, $k_{3}=7$ and different choices of $k_{24}$.}
 \label{fig:a_k24}
\end{figure}

As shown in Figures~\ref{fig:a_eta}, \ref{fig:a_k3} and \ref{fig:a_k24}, when Ericksen's inequalities hold, $\amin$ is a concave function, nearly linear for small values of $\sigma$, attaining in general its maximum at some $\sigma_{M}< 1$. The value of $\sigma_{M}$ depends on the choice of $\eta$, $k_{3}$, and $k_{24}$: it decreases as $\eta$ or $k_{3}$ increase (see Figs.~\ref{fig:a_eta}, \ref{fig:a_k3}), and as $k_{24}$ decreases (see Fig.~\ref{fig:a_k24}). As a consequence, the corresponding function $\almin$ (obtained from Equation~\eqref{eq:divergence_free_alpha}), which represents the twist angle in the toroidal frame $\torframe$, can attain its maximum $\maxtwist$ \emph{within} the torus $\body$.

\begin{figure}[h]
\centering
  \includegraphics[width=0.3\textwidth]{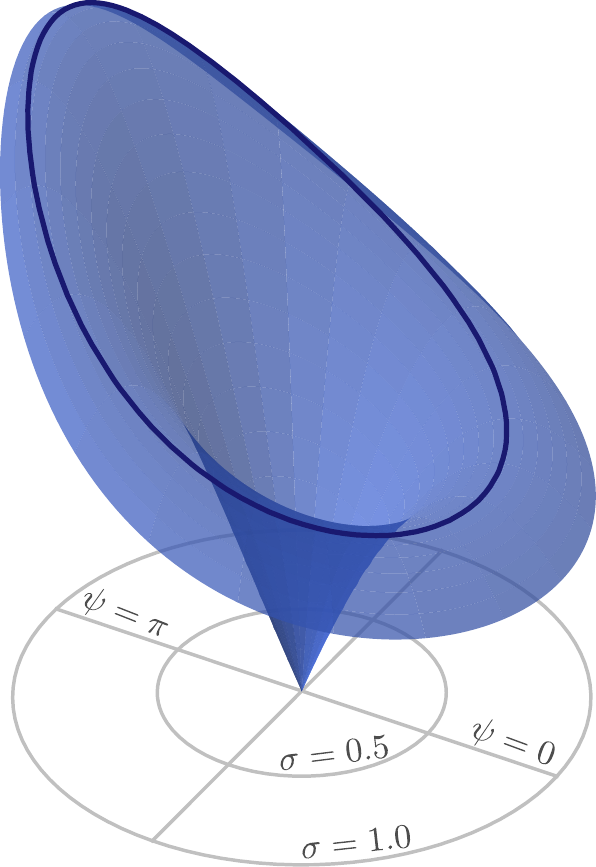}
  \caption{(Colour online) Lily-shaped surface that represents the plot of $\almin$ corresponding to  $\amin$ through Equation~\eqref{eq:divergence_free_alpha}. The dark (blue) line describes the maximum twist deflection  $\alpha$ along every fixed radial direction in the cross-section of the torus. In this experiment: $\eta=0.5$, $k_{3}=8.5$, and $k_{24}=0.1$.}
  \label{fig:alpha}
\end{figure}

\begin{figure}[h]
\centering
  \includegraphics[width=0.2\textwidth]{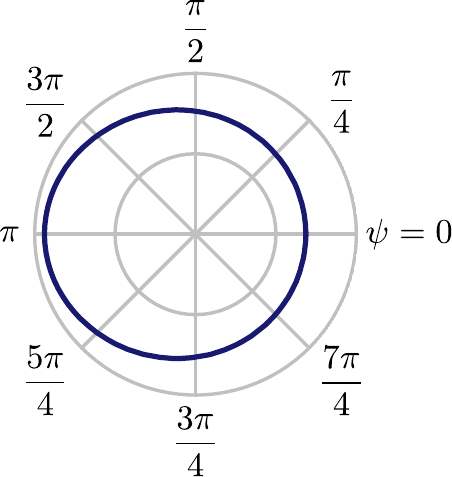}
  \caption{(Colour online) The line of radial maximum twist, that is, the projection on the cross-section of the torus $\body$ of the dark (blue) line of Figure~\ref{fig:alpha}, for $\eta=0.5$, $k_{3}=8.5$ and $k_{24}=0.1$.}
  \label{fig:argmax}
\end{figure}

An example in which $\maxtwist$ is located in the interior of $\body$ is provided by choosing $\eta=0.5$, $k_{3}=8.5$, and $k_{24}=0.1$, as in Figure~\ref{fig:alpha}, where a lily-shaped surface represents the twist angle $\almin(\sigma,\psi)$, for $\sigma$ and $\psi$ varying in the cross-section of the torus. The dark line marked on this surface describes how the maximum twist deflection along every fixed radial direction  varies with the angle $\psi$. The projection of this curve on the torus' circular cross-section designates the line of \emph{radial maximum twist} (rmt); it  encircles the centre of the cross-section, running through the whole interior of the torus, from the outer radius (at $\psi=0$) to the inner radius (at $\psi=\pi$), being closer to the latter end than to the former, see Figure~\ref{fig:argmax}. 
\begin{figure}
 \centering
 \includegraphics[width=0.9\linewidth]{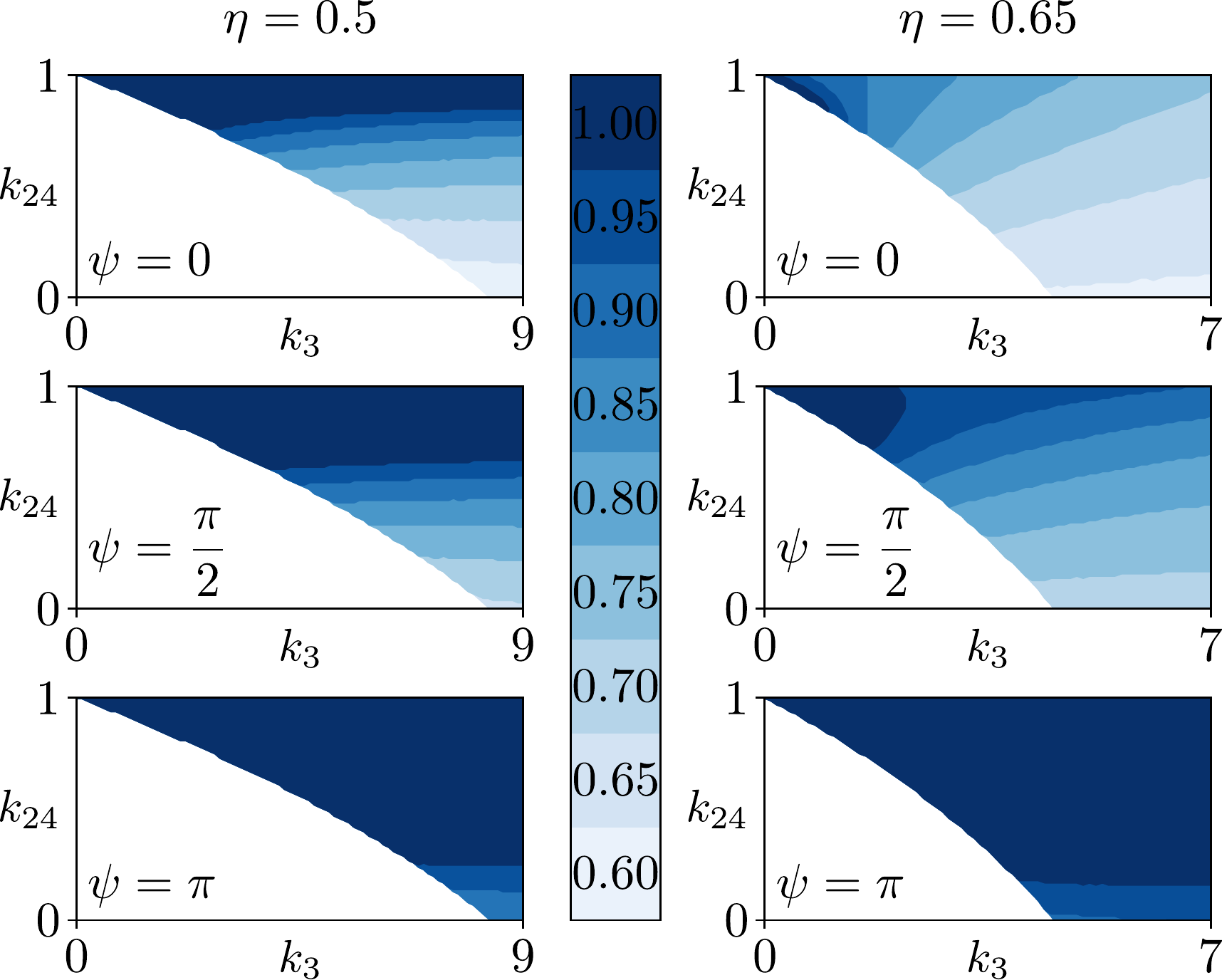}  
 \caption{(Colour online) Contour plots of $\rmt$ on the plane  $(k_{3},k_{24})$ of the scaled elastic constants, for $\eta=0.5$ (left) and $\eta=0.65$ (right) and for $\psi=0$ (top), $\psi=\dfrac{\pi}{2}$ (center) and $\psi=\pi$ (bottom).   The white bottom-left region in each panel is where the bend-only director field $\nbend$ is stable. Darker shades of blue correspond to higher values of $\rmt$, and so the line of radial maximum twist deflection is closer to the boundary of the torus.}
\label{fig:sigma_max}
\end{figure}

By increasing $\eta$ or $k_{24}$ the line rmt opens up and touches the boundary $\boundary$ on the inner side of the torus, generally remaining well inside $\body$ on the other side.  
In polar coordinates $(\psi,\sigma)$, the line rmt is represented by a function $\sigma=\rmt(\psi)$, which changes with  the geometric parameter $\eta$ and the scaled elastic constants $k_3$ and $k_{24}$. 
The contour graphs in Figure~\ref{fig:sigma_max} show how $\rmt$ depends on $(k_{3},k_{24})$, for $\eta=0.5$ and $\eta=0.65$; for $\psi$, we have chosen  its three most significant values. The farther is $\rmt$ from unity, the farther is the line rmt from  the boundary of the torus: visually, these cases correspond to the lighter shades of blue in Fig.~\ref{fig:sigma_max}, whereas the the darker shades  correspond to lines rmt closer to the boundary. Accordingly,  the white regions on the bottom-left corners of the plots in Fig.\ref{fig:sigma_max} represent values of  $(k_{3},k_{24})$ for which $\almin$ is zero and the bend-only director field $\nbend$ is stable. For a fixed $\eta$, the shades of blue darken uniformly in the plane $(k_{3}, k_{24})$ as the angle $\psi$ increases from $0$ to $\pi$, in perfect accord with Figure~\ref{fig:argmax}: the line rmt is off-centre, closer to the inner equator than to the outer equator.   The intermediate value $\psi=\dfrac{\pi}{2}$ is considered  here both for completeness and for being somewhat special, as there, by Equation~\eqref{eq:divergence_free_alpha}, $\almin=\arcsin\amin$. The countour plots of Figure~\ref{fig:sigma_max} also reveal that for high values of $k_{24}$ the overall maximum twist deflection $\alpha_\mathrm{M}$  tends to  fall on the boundary $\boundary$, suggesting that the inner twist is surface-driven. However, this tendency is significantly reduced upon increasing $\eta$: then  the inner twist would be surface-driven only for very small values of $k_{3}$.

\begin{figure}[h]
\centering
  \includegraphics[width=0.5\linewidth]{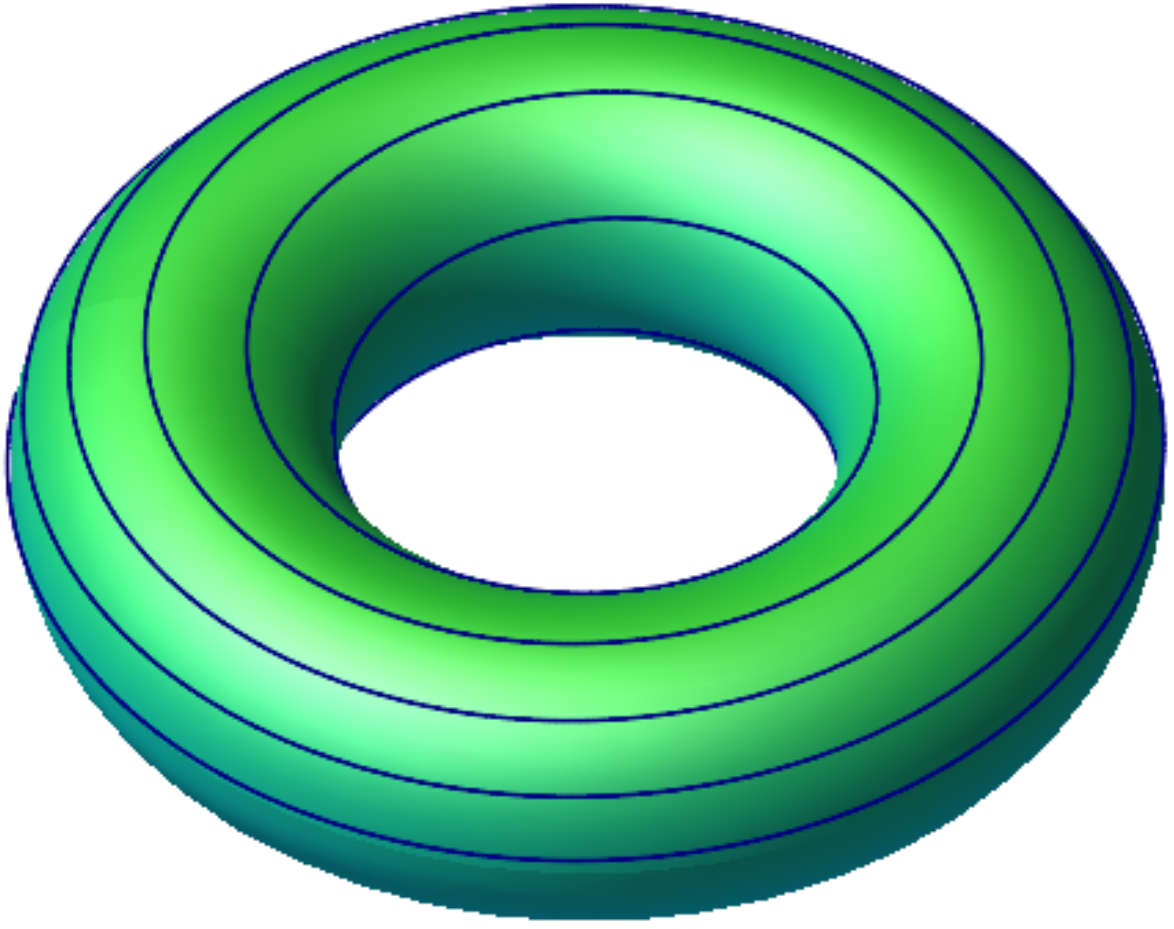}
  \caption{(Colour online) Vector lines of the director field $\n$ for the same $\almin$ as in Figure \ref{fig:alpha} ($\eta=0.5$, $k_{3}=8.5$, and $k_{24}=0.1$), wrapped around the internal toroidal shell at $\sigma=0.8$.}
  \label{fig:integral_lines}
\end{figure}

Finally, by both Equations~\eqref{eq:n_alpha_representation} and~\eqref{eq:divergence_free_alpha}, the knowledge of $\amin$ also allows us to describe  the vector field $\n$ that minimizes $\Free$.  Figure~\ref{fig:integral_lines} shows the vector lines of $\n$ for $\eta=0.5$, $k_{3}=8.5$ and $k_{24}=0.1$ (same parameters as in Figure~\ref{fig:alpha}) on the internal toroidal shell at $\sigma=0.8$: they approach  the parallels on the outer side and bend slightly towards the meridians  on the inner side.

\section{Validation}\label{sec:validation}

As a validation of the optimization method presented in Section \ref{sec:deep_learning} we tested it on the very special case $\eta\approx0$. For small values of $\eta$, indeed, the torus can locally be approximated by a cylinder, and so we can compare $\amin$ with the analytical solution $\acyl$, existing for $k_{24}>1$ only, given by Davidson et al.\cite{davidson:chiral}
\begin{equation}\label{eq:a_cylinder}
\acyl(\sigma):=\sin\left(\arctan\dfrac{2\sqrt{k_{24}(k_{24}-1)}\sigma}{\sqrt{k_{3}}\sqrt{k_{24}-(k_{24}-1)\sigma^{2}}}\right)\,.
\end{equation}
\begin{figure}	
 \centering
 \includegraphics[width=0.9\linewidth]{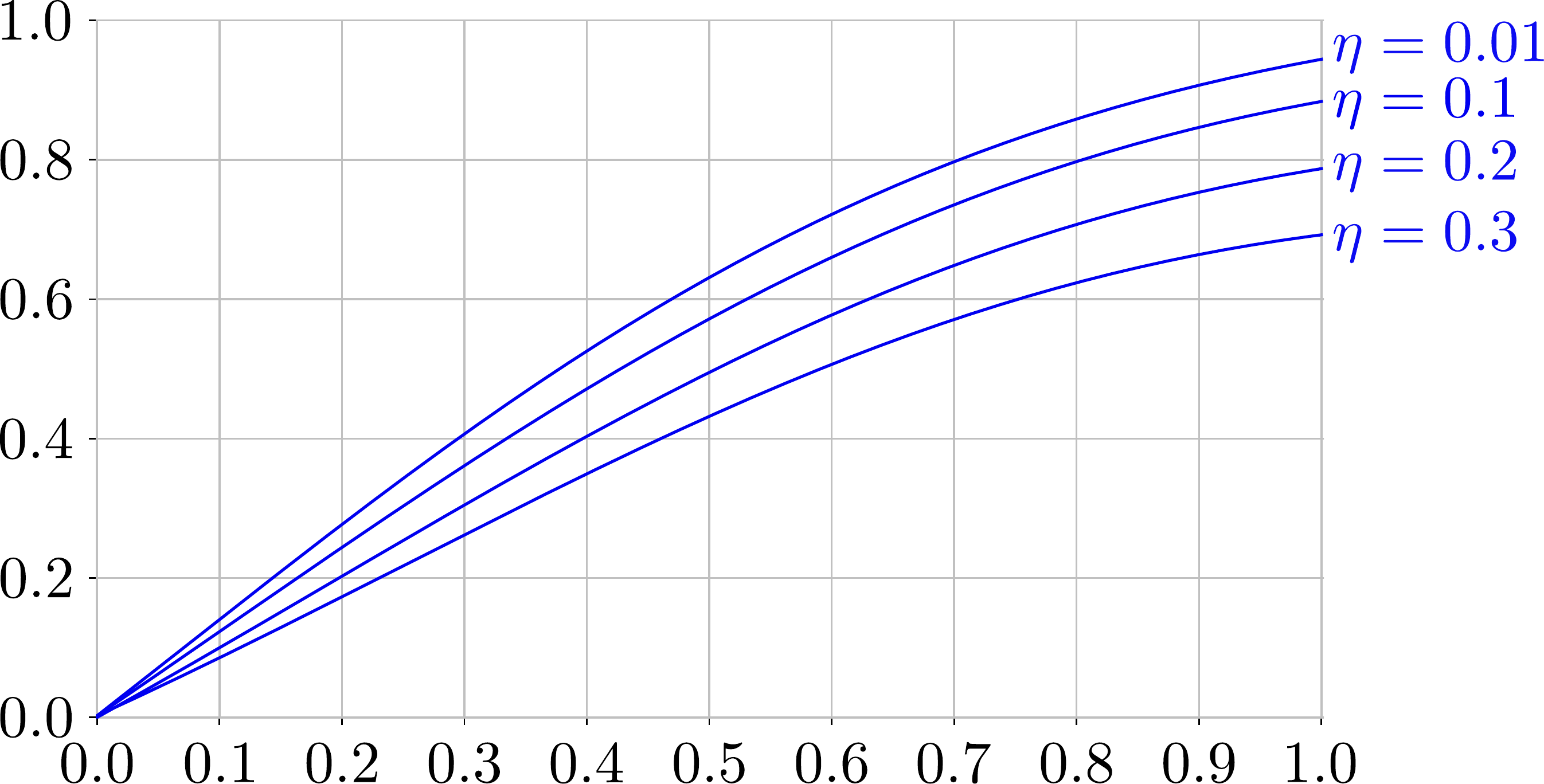}
 \caption{(Colour online) In the limit $\eta\to 0$ the torus becomes a cylinder and $\amin$ converges to the expected $\acyl$ in Equation~\eqref{eq:a_cylinder}. Here is an example for $k_{3}=1$ and $k_{24}=2$: for $\eta=0.01$ the error $\|\acyl-\amin\|_{\infty}$ is $\mathtt{\sim}10^{-3}$.} 
 \label{fgr:cilinder_limit_a}
\end{figure}
\begin{figure}	
 \centering
 \includegraphics[width=0.9\linewidth]{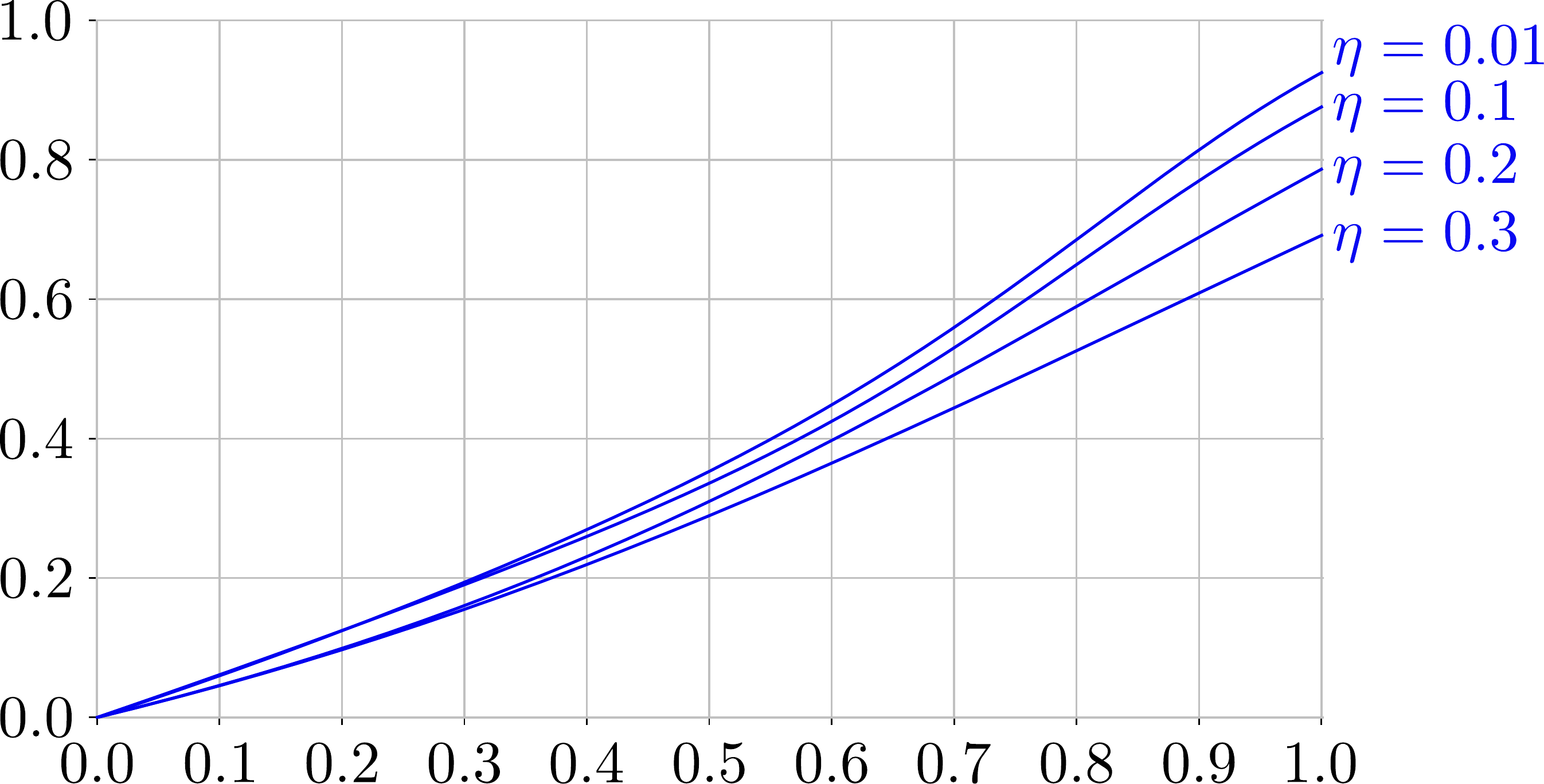}
 \caption{(Colour online) In the limit $\eta\to 0$ the torus becomes a cylinder and $\amin$ converges to the expected $\acyl$ in Equation~\eqref{eq:a_cylinder}. Here is another example for $k_{3}=8$ and $k_{24}=4$: for $\eta=0.01$ the error $\|\acyl-\amin\|_{\infty}$ is $\mathtt{\sim}10^{-3}$.} 
 \label{fgr:cilinder_limit_b}
\end{figure}

Figures~\ref{fgr:cilinder_limit_a} and~\ref{fgr:cilinder_limit_b} show the results of two series of experiments for $\eta\to0$: the former with $k_{3}=1$ and $k_{24}=2$, and the latter with $k_{3}=8$ and $k_{24}=4$. For $\eta=0.01$, the error $\|\acyl-\amin\|_{\infty}$ (where $\|\cdot\|_{\infty}$ denotes the standard sup norm in $[0,1]$) is negligible in both cases.
\begin{figure}
 \centering
 \begin{subfigure}{0.5\textwidth}
  \centering
  \includegraphics[width=0.8\linewidth]{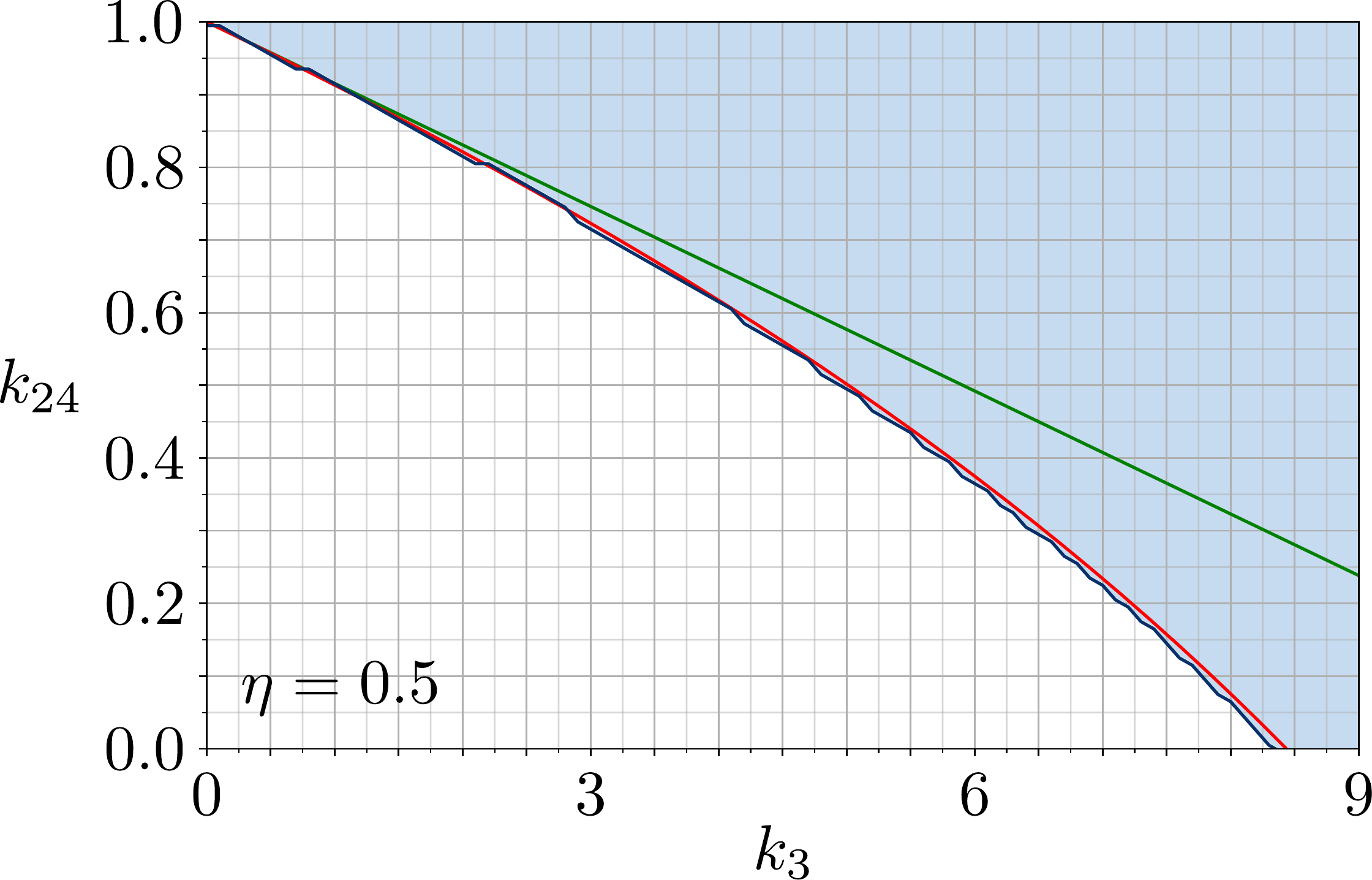}
 \end{subfigure}
 \begin{subfigure}{0.5\textwidth}
  \centering
  \includegraphics[width=0.8\linewidth]{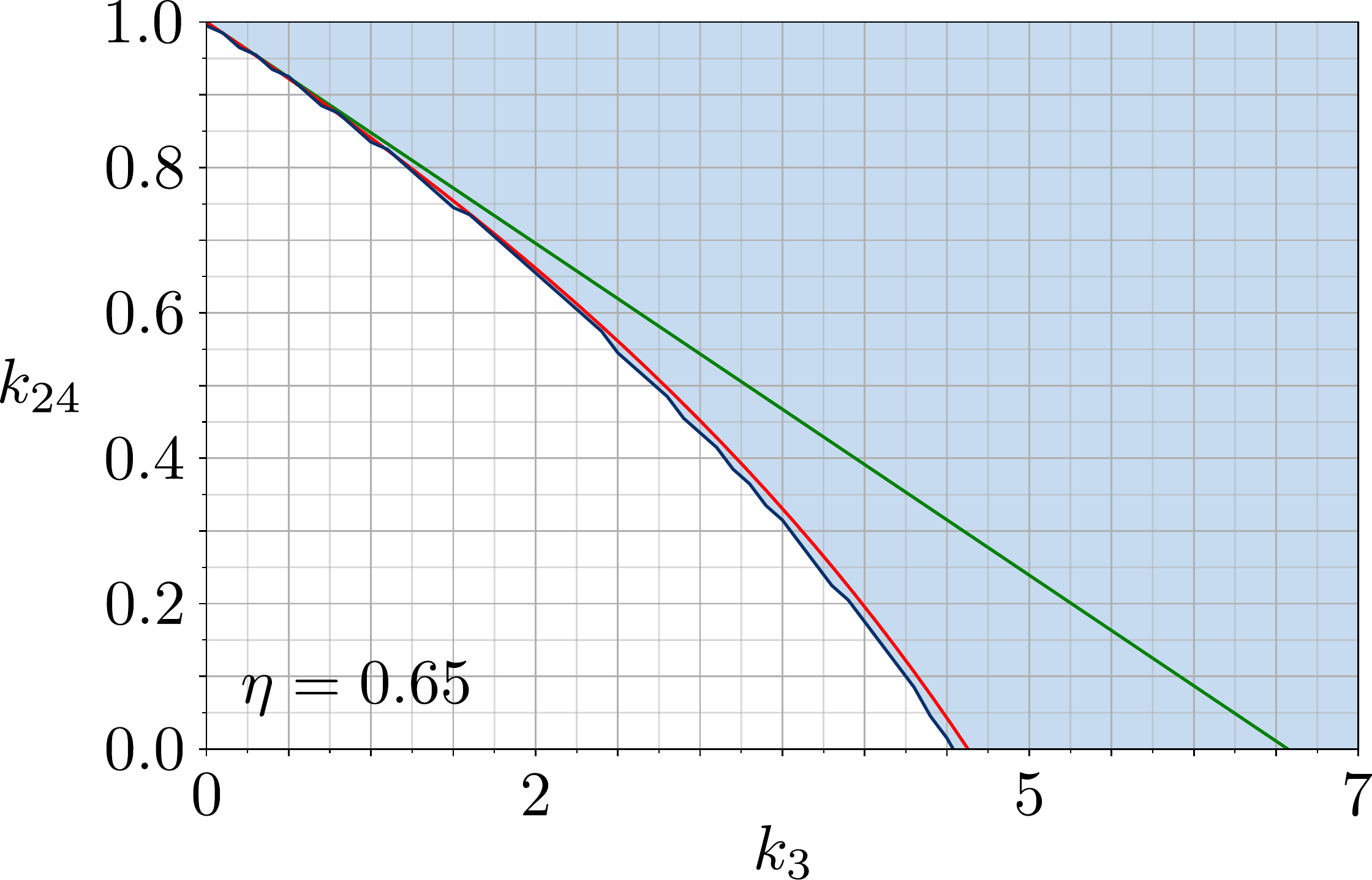}
 \end{subfigure}
 \begin{subfigure}{0.5\textwidth}
  \centering
  \includegraphics[width=0.8\linewidth]{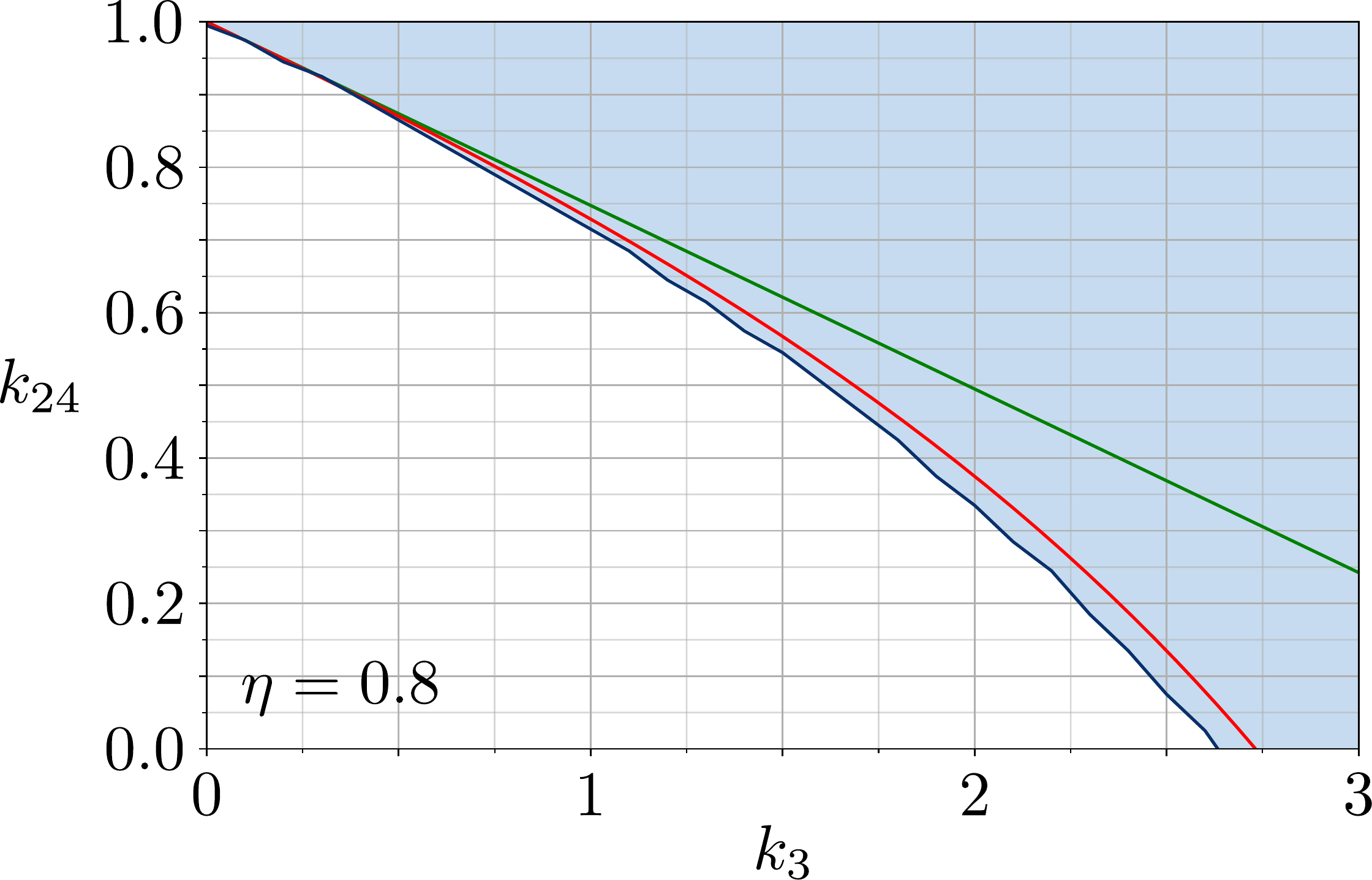}
 \end{subfigure}
 \caption{(Colour online) Comparison between instability conditions for $\eta=0.5$ (top), $\eta=0.65$ (center) and $\eta=0.8$ (bottom). The green straight line delimits  the instability region found  by Koning et al.\cite{koning:saddle-splay}, while the red line delimits the instability region obtained by Pedrini and Virga\cite{pedrini:instability}. The light blue area represents the refined instability region found by performing a grid search in the plane $(k_{3}, k_{24})$ with strides $10^{-1}$ and $10^{-2}$, respectively, and by marking as instable any point for which $\|\amin\|_{\infty}> 10^{-3}$, where $\amin$ is the result of the optimization method described in Section~\ref{sec:deep_learning}.}
 \label{fgr:instability_conditions}
\end{figure}

A further validation of our method is shown in Figure~\ref{fgr:instability_conditions}. Here, we compare the instability region for the pure-bend director field $\nbend$ found numerically  by our optimization method with the instability conditions  given analytically by Koning et al.\cite{koning:saddle-splay} and by Pedrini and Virga\cite{pedrini:instability}, with different choices of admissible test functions.  For $\eta=0.5$, $\eta=0.65$, and $\eta=0,8$, we performed a grid search in the plane $(k_{3},k_{24})$ with strides $10^{-1}$ in $k_{3}$ and $10^{-2}$ in $k_{24}$. Each point of the grid is marked as \emph{unstable} whenever the corresponding $\amin$ has $\|\amin\|_{\infty}> 10^{-3}$. For all  values of $\eta$, the instability region obtained in this way is an improvement on both  regions predicted analytically. The graphs in Fig.~\ref{fgr:instability_conditions} also show that the  instability analysis of Pedrini and Virga\cite{pedrini:instability} was quite on the mark: it effectively provides a reliable stability criterion. 

Finally, our optimization method proves effective even if less guided. Further experiments showed that suppressing the boundary condition imposed by Equation~\eqref{eq:minimizer_boundary} (thus leaving the representation in Equation~\eqref{eq:representation} unconstrained) does not significatively change the final result: $\amin$ remains essentially the same function, still satisfying $\amin(0)=0$ to a very good approximation.

\section{Conclusions}\label{sec:conclusions}
We  studied toroidal nematics subject to planar degenerate anchoring for the director by solving numerically the fully non-linear problem for the minimisers of Frank's elastic free energy.
When the pure-bend director field $\nbend$ (with vector lines  along the parallels of all internal toroidal shells) becomes unstable, two twisted director fields (with opposite handiness) take its place. It is quite natural to expect that such an instability is surface-driven, and so the maximum twist injected in the director texture should appear at the boundary of the torus. Remarkably, however, this turned out \emph{not} to be the case. Typically, the twist distribution inside the torus is instead well described by a lily-like surface. The radial maximum twist deflection takes place along an inner, off-centre ring of the torus' circular cross-section, away from which the twist angle (radially) decreases in both directions: to zero on the centre of the cross-section and to a positive value on the cross-section's periphery. We think that such a characteristic twist structure could be observed experimentally and might become the optical  signature of toroidal nematics.

The elastic free-energy functional that describes toroidal nematics in the fully non-linear case is rather involved, especially when all elastic constants are considered as free parameters. To cope with such a complexity, we developed an \emph{ad hoc} deep-learning optimization method that proved quite versatile and also reliable, in light of two independent validation tests, in which the outcomes of our method were confronted with analytical predictions available in the literature. We plan to exploit further this method, which is duly documented  and made available on the web, in the study of similar problems, among which is primarily the stability analysis of toroidal nematics subject to strong anchoring conditions for the director. Our objective will be to show under what condition on the smallness of the twist elastic constant $K_2$ the characteristic lily-like distribution of twist found here also arises there (if it ever does). We would expect that the answer to this question could be significant for chromonic liquid crystals,\cite{masters:chromonic} for which $K_2$ is by far the smallest among all four elastic constants.

\section*{Conflicts of interest}
There are no conflicts to declare.

\section*{Acknowledgements}
The work of A.P. has been supported by the University of Pavia under the FRG initiative, meant to foster research among young postdoctoral fellows.

\appendix
\section{Deep-learning code}\label{sec:appendix}

\balance

The code for the optimization method is available at the following repository:
\url{https://bitbucket.org/AndreaPedriniUniPV/lily-like_twist_distribution_in_toroidal_nematics/}

The package contains the main file \texttt{run.py} and a \texttt{lib} folder with all the ausiliary code; the code is written in Python\footnote[4]{Python Software Foundation. Python Language Reference, version 3.x. Available at \url{http://www.python.org}} and all further requirements are listed in \texttt{requirements.txt}.

A \texttt{README.md} file, also available on the web, contains all information and instructions needed to run properly  the code, to change the values of parameters $\eta$, $k_{3}$, and $k_{24}$, and to modify some advanced configurations of the program. It also describes four examples, chosen from the experiments run to produce the figures presented in the main body of the paper.

\providecommand*{\mcitethebibliography}{\thebibliography}
\csname @ifundefined\endcsname{endmcitethebibliography}
{\let\endmcitethebibliography\endthebibliography}{}

\end{document}